# Optimal Voltage for Nanoparticle Detection with Thin Nanopores

Yinghua Qiu*

Department of Physics, Northeastern University, Boston, 02115, MA

y.qiu@neu.edu

## Abstract:

The resistive-pulse technique provides a fast and label-free method for nanoparticle detection. In order to achieve a higher sensitivity, thin nanopores, such as silicon nitride pores, are usually considered. In this paper, nanoparticle detection has been mimicked with simulations. We found the surface charges of the particle can affect the current blockade obviously in short pores, especially under high electric fields. For particles with a surface charge density higher than $-0.02$ C/m$^2$, its current blockade ratio depends on the applied voltage closely. From our simulation results, an optimal voltage can be found for the particle detection, under which the current blockade ratio doesn't depend on the surface charge density of the particle. This optimal voltage was obtained by the balance of current increase and decrease caused by cations and anions, respectively, due to the negative surface charges of particles. From the systematical study, the optimal voltage was found to work like a property of the system which only depends on the electrolyte type. We think our finding can provide some help to the accurate particle detection in experiments.

**TOC**

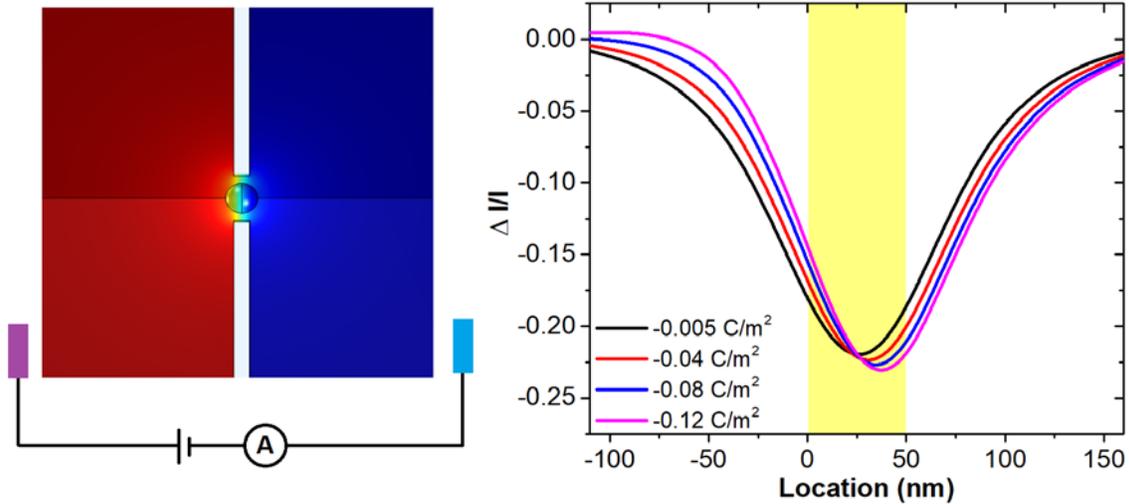

**Introduction:**

Fast and label-free nanoparticle detection has become of great importance due to its application in virus sensing, disease diagnosis, and drug delivery.[1, 2] Resistive-pulse technique originated from Coulter counter[3] provides a perfect candidate for nanoparticle detection because of its advantages, such as individual particle detection, easy operation, as well as high accuracy and throughput. Since its invention in the 1950s, resistive-pulse technique has been used for various objects detection, from nanoscale biomolecules,[4-6] such as DNA, RNA and viruses, to microparticles[7, 8] like cells and bacteria.

For example, exosomes as a kind of nanovesicles with a size ranging from 30 to 150 nm[9] have been investigated extensively recently, due to their role in cell communication and cancer diagnosis.[10] Some results show that the size of the exosomes may have some relationship with the human health status.[11] It's important to find an easy way to detect the size of exosomes precisely. Atomic force microscope (AFM)[11] and transmission electron microscope (TEM)[12] have been used to detect the size of exosomes. However, the efficiency of both methods is very low. Due to the size range of exosomes, traditional flow cytometry cannot give accurate detection.[12] Nanoparticle tracking analysis (NTA) can provide particle detection as small as ~50 nm. However, fluorescence labeling is usually needed.[13] In this case, accurate exosome detection with resistive-pulse technique becomes promising.

In a resistive-pulse experiment, electrokinetic or pressure driven particles will cause a transient current change, called a resistive pulse, when it passes through a pore due to the increase of the system resistance.[14, 15] Based on the dependence of magnitudes of the resistive pulses on the size of the detected objects, the size information of the particle can be easily obtained through classical theories.[14]

In order to simplify the translocation process of bio-particles, like viruses, exosomes, or cells, hard artificial spheres are usually used to explore the fundamental physics of particle translocation through the pore because they are more easily prepared. In this paper, nanoparticles with a diameter from 50 to 125 nm have been considered.

During the resistive-pulse experiments, in order to get a good sensitivity, it's very important to find a pore with suitable size for the detection of specific particles.[14] For the pores with tens of micrometer in length, cylindrical pores are hard to detect the object less than 100 nm in size, because the transient change in the ionic current, which depends on the volume ratio of the object and the pore, is very tiny during its translocation.[14, 16] Some groups have tried to detect exosomes by huge conical pores, which cannot give a high sensitivity.[17, 18] For the nanoparticles with ~100 nm in diameter, silicon nitride pores could provide a great tool for resistive-pulse detection.[19, 20] With nanofabrication technique, silicon nitride chips with different thicknesses can be easily prepared through chemical vapor deposition (CVD). Then, nanopores can be fabricated by focused ion beam (FIB) drilling[21] or dielectric breakdown.[22-24] FIB can be used for thicker membranes above ~30 nm to micrometers,[20] and dielectric breakdown provides an easy way to make nanopores with controlled sizes for the membranes with ~30 nm or less in thickness.[24] Based on the advanced data acquisition technology, the translocation events of particles through thin nanopores can be easily obtained even with durations as short as only several microseconds.[25, 26]

At solid-liquid interfaces, surface charges usually appear due to many mechanisms, such as deprotonation and ion adsorption.[27] From earlier investigations, surface charge density, as an important property of nanoparticles, can affect the resistive-pulse detection obviously, not only the duration time[28-30] but also the current blockade.[31-34] With microscale polymer pores, Qiu et al.[31] found that strongly charged polystyrene particles with 400 nm in diameter can cause much higher current blockade due to the appearance of ionic concentration polarization across the particle.[35] When the particle moves through the pore, a huge depletion zone of ions will form in the front of the particle which will enhance the current blockade. The same trend was found by Chen et al.[32] with simulations of nanoparticles using SiN pores. With microscale glass conical pores, Lan et al.[33] found that particles with different surface charge densities can cause different magnitudes of resistive pulses. Different from the results of Qiu et al., particles with higher surface charge density can cause a lower current blockade due to the accumulation of ions which eventually reduces a biphasic pulse shape.[28, 36, 37] From simulations, Wang et al.[34] also obtained results with the similar trend to that of Lan et al. using thin nanopores. This may be due to the low voltages used in their systems.

For the detection of electrokeitic driven particles, the voltage across the nanopore is an important parameter which has been seldom considered. The selection of voltage in the experiment is usually arbitrary. Because of the short length of the nanopore, high electric field strength can be easily formed by applying a weak voltage. Surface charges may cause obvious ionic concentration polarization across the particle, which will cause higher current blockade. Then, it will be difficult to predict the particle size based on the magnitude of pulses. Based on the earlier studies,[31, 32] aqueous solutions with a high concentration may solve this problem because more counterions can screen the particle surface charges better. However, high concentrations can also introduce problems like particle aggregation.[27, 38] In this paper, COMSOL simulations have been used to mimic the detection of ~100 nm nanoparticles with SiN pores. For the particles with the same size but different surface charge density, an optimal voltage for the current blockade detection was found under which the magnitude of the resistive pulse doesn't depend on the surface charge density of the particle. Above this voltage, current blockade becomes deeper for higher surface charge density.[31, 32] While, below this voltage, current blockade gets lower for higher surface charge density which corresponds to the results obtained by Lan et al.[33] and Wang et al.[34] Our finding has given a whole picture of the influence of voltage and surface charge density on the current blockade, which can provide some help for the experimental detection with resistive-pulse technique.

Please note that in this work, we haven't considered the effect of particle trajectories[39, 40] and pore shape[41] on the current blockade due to the huge consumption of the computing time.

**Simulation Details:**

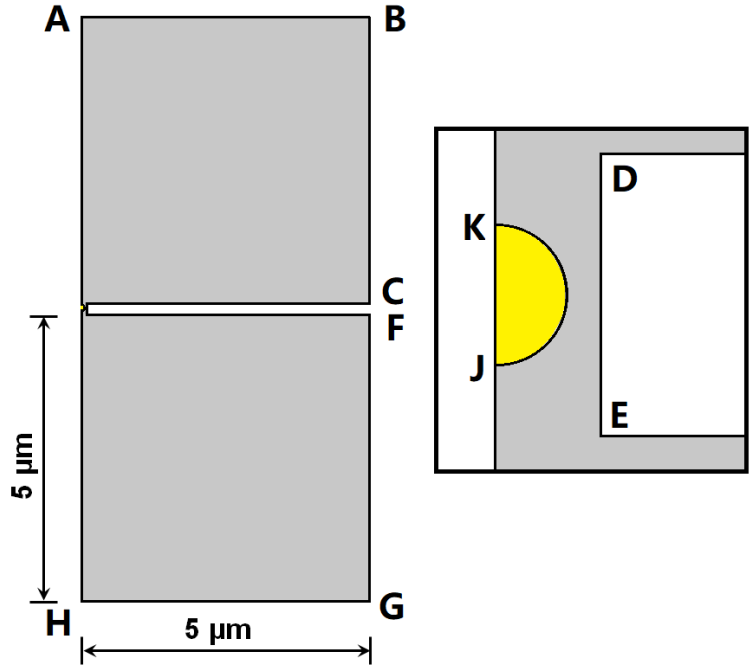

Figure 1 Scheme of the simulation. Zoomed-in pore region is shown on the right. Yellow part shows the particle.

3D simulation was conducted by solving coupled Poisson-Nernst-Planck (PNP) and Navier−Stokes (NS) equations with COMSOL Multiphysics 5.2 package to model steady-state solutions for ionic current at room temperature (298 K).[41] Figure 1 shows the scheme of the simulation system. Table 1 lists the boundary conditions used in this model. The length of the cylindrical pore ranges from 10 to 500 nm, and its diameter can be 150, 175, 200 and 250 nm. –0.005 C/m$^2$ was selected as the surface charge density of SiN membrane.[32, 42] For the inner surface of the pore, the mesh size of 0.1 nm was used to consider the effect of electrical double layers. For the charged boundaries of the reservoirs the mesh of 0.5 nm was chosen to lower the memory cost during calculation.[41] For the nanoparticle, 50, 75, 100 and 125 nm were selected for the diameters. The surface charge density of the particle was set from –0.005 to –0.12 C/m$^2$. As predicted by Grahame equation,[27] –0.1 C/m$^2$ can give a particle a surface potential around –89 mV in 0.1 M solution which will be higher in solutions with lower concentrations. 0.1 nm mesh size was chosen for the particle surfaces. An example of mesh plot was shown in Figure S1. In this paper, most simulations were performed in KCl aqueous solution. The dielectric constant of water was assuming as 80. Diffusion coefficients for potassium and chloride ions were assumed equal to the bulk value of $1.92 \times 10^{-9}$ m$^2$/s and $2.03 \times 10^{-9}$ m$^2$/s, respectively.[40] The salt concentration was 0.1 M in most cases. 0.01, 0.02, 0.05, 0.2 and 0.3 M were used to consider the effect of ionic concentration. NaCl and LiCl solutions were also used to consider the ionic mobility influence. The ionic mobility for Na$^+$ and Li$^+$ ions were set as $1.33 \times 10^{-9}$ m$^2$/s and $1.03 \times 10^{-9}$ m$^2$/s, respectively.[43] Voltages ranging from 5 to 600 mV were applied in each case of the simulations.

**Table 1**. Boundary conditions used in numerical modeling. Coupled Poisson-Nernst-Planck and Navier-Stokes equations were solved with COMSOL Multiphysics 5.2 package.

| Surface | Poisson | Nernst-Planck | Navier-Stokes |
|---|---|---|---|
| **AB** | constant potential $\phi=0$ | constant concentration $c_i=C_{bulk}$ | constant pressure $p=0$ no viscous stress $\mathbf{n}\cdot[\mu(\nabla\mathbf{u}+(\nabla\mathbf{u})^T)]=0$ |
| **BC, FG** | no charge $-\mathbf{n}\cdot(\varepsilon\nabla\phi)=0$ | no flux $\mathbf{n}\cdot=0$ | no slip |
| **CD, DE, EF** | $-\mathbf{n}\cdot(\varepsilon\nabla\phi)=\sigma_w$ | no flux $\mathbf{n}\cdot\mathbf{N}_i=0$ | no slip $\mathbf{u}=0$ |
| **HG** | constant potential $\phi=V_{app}$ | constant concentration $c_i=C_{bulk}$ | constant pressure $p=0$ no viscous stress $\mathbf{n}\cdot[\mu(\nabla\mathbf{u}+(\nabla\mathbf{u})^T)]=0$ |
| **AK, HJ** | axial symmetry | axial symmetry | axial symmetry |
| **curve JK** | $-\mathbf{n}\cdot(\varepsilon\nabla\phi)=\sigma_p$ | $\mathbf{n}\cdot\mathbf{N}_i=0$ | no slip $\mathbf{u}=0$ |

$\phi$, $\varepsilon$, $C_{bulk}$, $p$, $\mathbf{n}$, $\mathbf{N}_i$, $\mathbf{u}$, $\sigma_w$, $\sigma_p$, $V_{app}$, $\mu$ are the surface potential, dielectric constant, bulk concentration, pressure, normal vector, flux of ions, fluid velocity, surface charge density of the pore wall, surface charge density of the particle surface, applied voltage, and solution viscosity, respectively.

**Results and Discussions:**

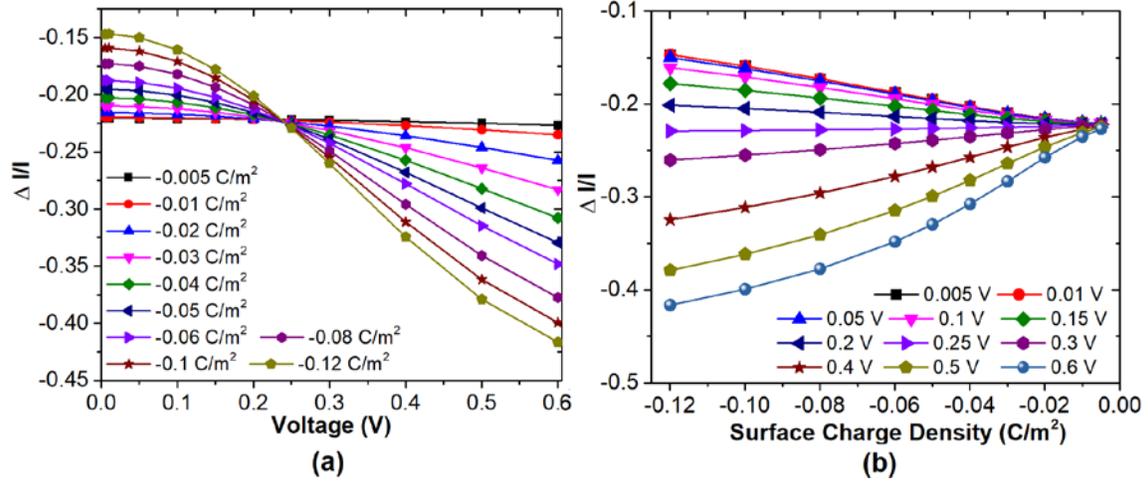

Figure 2. Current blockade ratios obtained from differently charged particles under different voltages. (a) Applied voltage as the variable, and (b) Surface charge density as the variable. Particle size is 100 nm in diameter. The pore is 150 nm in diameter and 50 nm in length. 0.1 M KCl was selected as the solution.

Nanoparticles with 100 nm in diameter but different surface charge densities have been tested using a pore with 150 nm in diameter and 50 nm in length. From the simulation, we can get the current values through the open pore and blocked pore as $I_o$ and $I_b$ by integration the ionic flux through the boundary AB or HG in Figure 1. For the open pore cases, the particle was put at 2 μm away from the pore. The current blockade i.e. the current change can be calculated as $\Delta I = I_b - I_o$. The current blockade ratio was calculated as $\Delta I/I_b$[14, 16], which is shown in Figure 2. To mimic SiN nanopores, −0.005 C/m$^2$ was selected as the surface charge density of the nanopore.[32, 42] In this paper, when the nanopore is selected, the baseline from the pore is determined. Then, if the current blockades are the same, the blockade ratios are also the same. During the simulations, the particle was set in the center of the nanopore. From the results, when the particle has a lower surface charge density such as −0.005 C/m$^2$, the current blockade ratio doesn't depend on the voltage applied across the pore. While, as the surface charge density of particles increases to −0.02 C/m$^2$, the dependence of current blockade ratio on the applied voltage becomes more obvious. For example, when the particle is charged at −0.03 C/m$^2$, which has a surface potential as around −38 mV,[27] the current blockade ratio at 0.6 V is ~35% higher than that from 0.1 V. This means the particle size detected at 0.6 V is much larger than that at 0.1 V based on the classical theories.[20, 34]

From Figure 2(a), it's very easy to find a voltage under which all the 100 nm particles with different surface charge densities have the same current blockade ratio. Below this voltage, like at 0.1 V, a lower current blockade ratio is caused by a particle with a higher surface charge density as found in the earlier work.[33, 34] While, above this voltage, such as at 0.4 V, the particle with a higher surface charge density has a higher current blockade ratio than that of a particle with a lower surface charge density.[31] The same results are also shown as the current blockade ratio with the surface charge density at the same electric field in Figure 2(b). With a lower voltage across the pore, the current blockade ratio gets lower with the surface charge density increasing. This is due to the enhanced ionic concentration near the particles in the pore. Under a higher voltage, the current blockade ratio enhances with the surface charge density. In a strong electric field, obvious ionic concentration polarization across the particles can appear due to the fast movement of cations on the negatively charged particle surface (Figure S2).[29, 35]

Please note that in earlier investigation of dynamic simulations, concentration polarization in conical pores may need time to reach equilibrium in diluted solutions.[44] In order to confirm our stationary simulation here could provide reasonable results for the resistive-pulse detection of particles in short pores, we have conducted a time-dependent simulation as shown in Figure S3 and S4. Weak ionic concentration polarization across the particle appeared at the beginning of the simulation. Then, the concentration polarization became more obvious and reached equilibrium at ~3 μs. From the obtained dynamic current through the nanopore, under 0.1 and 0.25 V the total ionic current can reach equilibrium very fast within 5 and 9 μs. Please note that the current value at 0.1 μs is very close to the current at 10 μs, within a 2% difference. We think the current may depend on the net value of current increase caused by cations and the current decrease caused by the coins. However, the rough translocation time of the particle through a 50-nm-in-length pore was evaluated as ~0.6 μs with Helmholtz-Smoluchowski equation (Supporting Information).[20, 45] This value of 0.6 μs may be not true, because the classical theory didn't consider the access resistance regions and the interaction between the particle and the pore, which will elongate the translocation time of a particle.[20] When the pore gets thinner, the access resistance dominates the total resistance of the system.[34] The main voltage drop happens in the access resistance regions. In real cases, the duration time for the particle translocation could be much longer. In the experimental work by Davenport et al.,[20] for the 100 nm particle with a zeta potential as –33.9 mV, its duration time is around 630 µs, when it passed through a SiN pore with a zeta potential as –44 mV, as well as 260 nm in diameter and 50 nm in length under an electric filed as ~3×10$^5$ V/m. Following their experiment results as well as the linear dependence of the particle speed on the electric field strength and the zeta potential of particle surfaces,[20, 45] the duration time for a particle translocation considered here can be ~18 µs, which is long enough for the concentration polarization to reach equilibrium. So, our stationary simulations in this work could be used to investigate the concentration polarization across the particle when it's passing through a thin nanopore.

In Figure 2(b), under a specific voltage ~0.25 V, the current blockade ratio of particles doesn't depend on its surface charge density which shows as a horizontal line. In the paper, we call this specific voltage as the optimal voltage. The optimal voltage can be picked in Figure 2(a) as ~0.237 V. This value is not exact. We picked the optimal voltage in the center of the overlap of the lines. Please note that this phenomenon doesn't depend on the consideration of surface charges on the reservoir walls. (Figure S5) But, it did depend on the consideration of electroosmotic flow in the pore. (Figure S6) We found that without the electroosmotic flow in the simulations, the increase of the surface charges on the particle surfaces can only make the blockade ratio decrease monotonously.

In order to explain the dependence of the current blockade ratio on the voltage and surface charge density, we investigated the current change caused by K$^+$ and Cl$^-$ ions, respectively, as shown in Figure 3. In this case, the current blockade ratio $\Delta I/I_b$ was calculated from only K$^+$ or Cl$^-$ ions. From our results, with the surface charge density increasing, more K$^+$ ions can be attracted to the particle surface which causes the increase of the K$^+$ ion current. At the same time, due to the electrostatic repulsion between the surface charges and Cl$^-$ ions, Cl$^-$ ions were depleted near the particle which is shown as the decrease of the Cl$^-$ ions current. From Figure 3, under lower voltages, the current increase of K$^+$ ions can overbalance the current decrease of Cl$^-$ ions which results in the lower current blockade ratio for more highly charged particles. When the voltage is above 0.25 V, strong electric field strength can cause an obvious concentration polarization which lowers the current increase of K$^+$ ions and enhances the current decrease of Cl$^-$ ions. Then, the final blockade becomes deeper with the voltage.

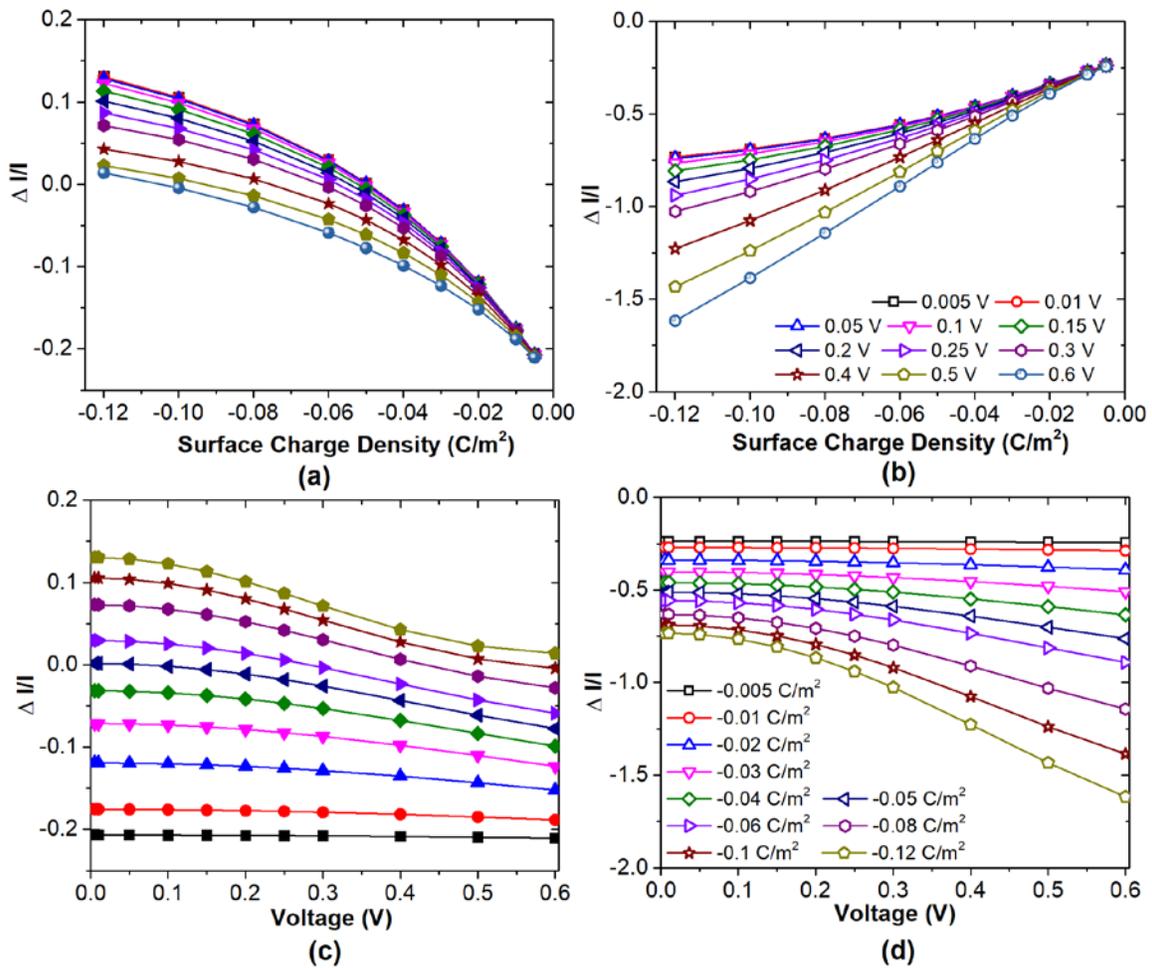

Figure 3. Current blockade ratio contributed by $K^+$ (a & c: shown as solid symbols) and $Cl^-$ (b & d: shown as open symbols) ions obtained from differently charged particles under different voltages. (a) and (b): Surface charge density as the variable. (c) and (d): Applied voltage as the variable. Particle size is 100 nm in diameter. The pore is 150 nm in diameter and 50 nm in length. 0.1 M KCl was selected as the solution.

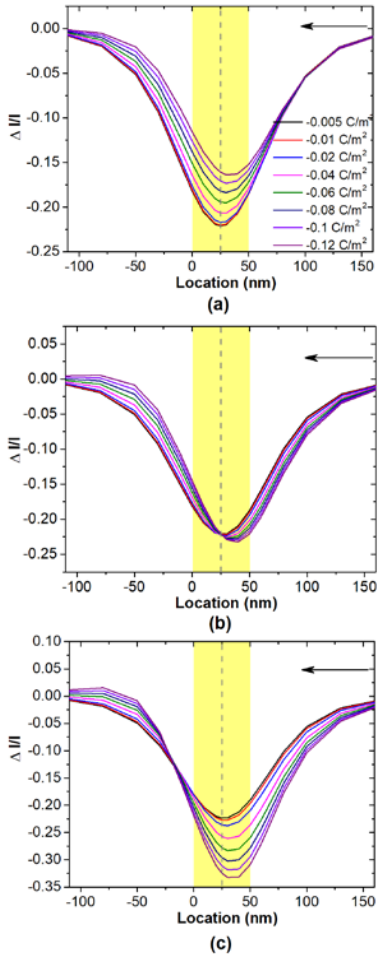

Figure 4. Current traces from 100 nm particles with different surface charge densities through a pore with 150 nm in diameter and 50 nm in length. (a) 0.1 V, (b) 0.237 V and (c) 0.4 V. The arrows show the electrophoresis direction of negatively charged particles. The pore region is shown as yellow. The dashed grey lines show the position of the pore center.

The current blockade traces from simulations with particles of different surface charge densities were obtained as shown in Figure 4. The particles were set at different locations along the pore axis to get the ionic currents. As predicted from Figure 2, the current blockade of particles doesn't depend on the surface charge density under 0.237 V as shown in Figure 4(b), which is especially good for the detection of highly charged particles.[16, 28] For the highly charged particles, we saw the weak current increase when the particle exits the pore.[19, 28, 37, 46] This was attributed to the enhanced local concentration of the counterions. Please note: as shown in Figure 4, with the surface charge density of the particle increasing, the position to get largest current blockade approaches to the entrance of the pore. This phenomenon is similar to the non-rectangular event shape obtained from a weakly charged particle in the micropore.[47] Due to the blockade of the charged particle, the cylindrical pore can be treated as a charged conical pore based on the differently confined space distribution along the pore axis. i.e. when the particle locates at the entrance of the pore, the entrance of the cylindrical pore can be treated as the tip of a conical pore. Under the electric field, the movement of $K^+$ ions from base side to tip side gives a lower current. In the opposite case, if the particle locates at the exit of the pore, the exit side can be treated as the tip of a conical pore. Then, the movement of $K^+$ ions is from tip side to base side which will give a higher current. Due to this phenomenon, it's not accurate to get the current blockade for more highly charged particles with the location at the center of the pore. Because the position of the deepest current blockade depends on the voltage and surface charge density, it will need much more calculation time to get the exact position for each case. Considering the highest value of the surface charge density of particles used in the earlier publication was 0.1 $C/m^2$,[32-34] in this paper, the highest surface charge density we discussed was set as -0.12 $C/m^2$ based on the almost same blockade value obtained in the pore center as the deepest current blockade. We also calculated the cases of more highly charged particles, such as -0.15, -0.2, -0.25, and -0.3 $C/m^2$, as shown in Figure S7. Under the optimal voltage, the current blockade ratio of the highly charged particles is very close to that of the weakly charged ones, within a maximum difference of 13%.

The optimal voltage has been explored using a series of pores with different lengths, diameters and surface charge densities. As shown in Figure 5 (a-c), the optimal voltage for the particle detection is ~0.225 V which is almost not influenced by the pore geometry, i.e. the pore diameter and length, or the surface charge density of the pore walls. While, the selection of stronger pore charges may affect the optimal voltage slightly: with the surface charge density of the pore increasing further, the optimal voltage decreases a little. As the surface charge density of the pore increases, the amount of counterions ($K^+$ ions) increases in the pore, which moves the balance of current increase caused by $K^+$ ions and current decrease caused by $Cl^-$ ions to the lower voltage side. Please note: we list all the current blockade ratio obtained from differently charged particles under different voltages for each case in the supporting information.

For the effect of the particle size of the optimal voltage, particles with 50, 75, 100 and 125 nm in diameter were considered. As shown in Figure 5(d), for the particle-pore diameter ratio is not very high, i.e. below 67%, the optimal voltage is not affected by the particle size. However, when the particle size is very large as 125 nm, a higher optimal voltage was obtained due to the enhanced attraction of the surface charges to the counterions and repulsion to the coions in more confined spaces.[45]

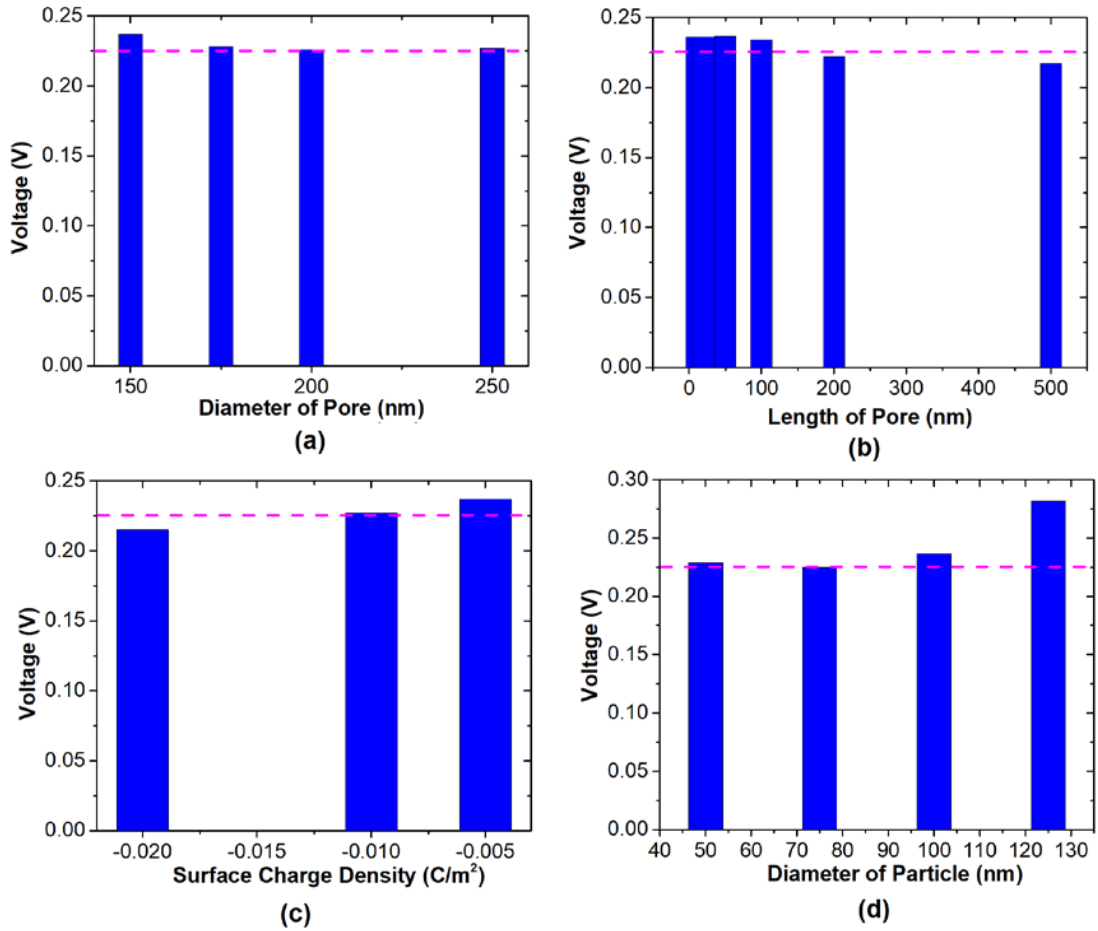

Figure 5. Optimal voltages obtained with particles from pores with different diameters (a), lengths (b), and surface charge densities (c). Pink dash lines show 0.225 V. Particle size is 100 nm in diameter. Pore is 50 nm in length (a, c), and 150 nm in diameter (b, c). (d) Optimal voltages obtained from particles with different sizes. The pore is 50 nm in length, and 150 nm in diameter. 0.1 M KCl was selected as the solution.

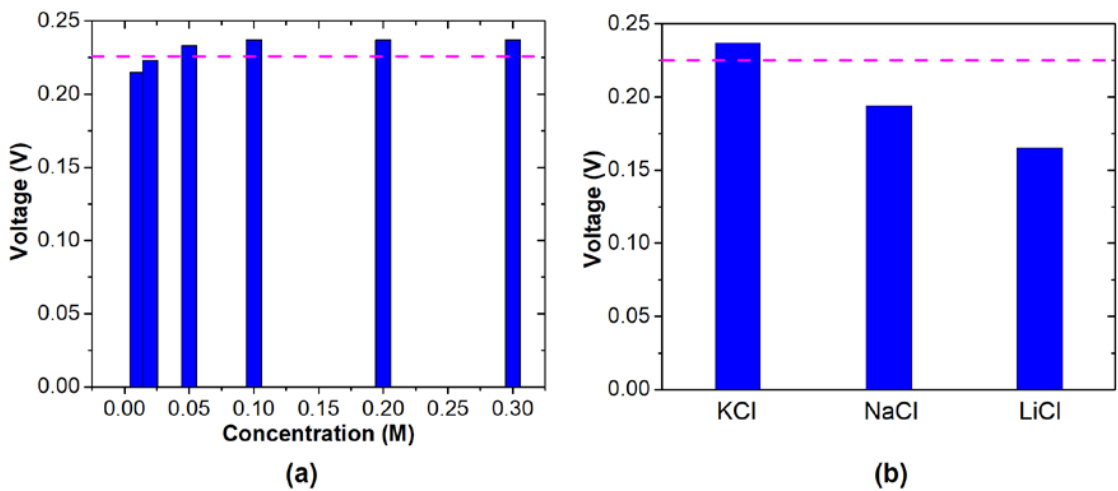

Figure 6. Optimal voltage obtained with particles in KCl solutions with different concentrations (a) and different types of electrolyte with 0.1 M (b). The pore is 50 nm in length, and 150 nm in diameter. Pink dash line shows 0.225 V.

Finally, the effects of ionic concentration and type on the optimal voltage were considered. As shown in Figure 6 (a), under different concentration of KCl solutions from 10 mM to 300 mM, the optimal voltage doesn't change obviously. With a higher concentration of solutions, the surface charge density effect on the current blockade becomes much less obvious (Figure S8) as the earlier reports.[31] In order to consider the effect of the ionic type on the optimal voltage, different monovalent solutions i.e. KCl, NaCl, and LiCl were used as 0.1 M in the simulations. We find that the mobility of the cation affects the optimal voltage obviously. For the cations with a lower ionic mobility, the current increase from cations caused by particle surface charges gets much weaker which will move the balance of the current increase from cations and current decrease from coions to the lower voltage side. We can also imagine that coions with a lower mobility will give a higher optimal voltage than that of coions with a higher mobility when the cations are the same.

## Conclusions:

Due to the tiny scale of nanopore and nanoparticles, the surface charge density of nanoparticles can affect the current blockade a lot in resistive-pulse detection. In order to exclude the surface charge effect on the current blockade of the particle, solutions with a high concentration are usually selected, which will also cause some problems like particle aggregation.[27] From our simulation of particle detection, an optimal voltage can be found in each case. Under the optimal voltage, the current blockade ratio of nanoparticle doesn't depend on the surface charge density of the particle anymore. The optimal voltage is not affected by the pore geometry, pore surface charge or solution concentration obviously, which is only influenced by the electrolyte type. In an experiment, when the pore and solution have been selected, the optimal voltage is determined. We think this finding will provide some help for the size detection of charged particles.

## Acknowledgment:

We'd like to thank Wei Zhang for her careful reading the manuscript.